\documentclass[twocolumn,preprintnumbers,amsmath,amssymb,pra]{revtex4}

\usepackage{graphicx,amsmath}
\usepackage{dcolumn}
\usepackage{bm}
\usepackage{amssymb}
\usepackage{epstopdf}
\usepackage{color}

\begin{document}

\title{Spontaneous emission of a three-level artificial atom in a one-dimensional open waveguide}

\begin{abstract}

We study the dynamical and spectral characteristics of a quantum three-level ladder system, interacting with a continuous electromagnetic field in one-dimensional open waveguide. Common realization of such systems is a waveguide QED setup - a superconducting artificial atom (transmon), coupled to an open microwave transmission line. We derive an analytical solution for spontaneous emission of initially excited atom, and use it to study the probability of state detection and spectral density of output photon states. We find that for strong coupling of transmon to a waveguide emitted photons show correlation in frequency and can have the same energies, even if the three-level system is anharmonic.

\end{abstract}

\date{\today}

\author{O. A. Chuikin}\email{ChuikinOA@yandex.ru}
\affiliation{Novosibirsk State Technical University, Novosibirsk, Russia}

\author{Ya. S. Greenberg}
\affiliation{Novosibirsk State Technical University, Novosibirsk, Russia}

\author{O. V. Kibis}
\affiliation{Novosibirsk State Technical University, Novosibirsk, Russia}

\maketitle

\section{Introduction}

Waveguide quantum electrodynamics is a modern and rapidly developing field of research that studies the interaction of atoms with photons in confined geometries of waveguides (see recent reviews \cite{Roy17} and \cite{Sheremet23}). It has a vast spectrum of applications in quantum optics and photonics, as well as in quantum technologies such as quantum communication and quantum information processing \cite{Gu17}. The adjacent field of cavity QED ~--- which usually focuses on manipulating quantum entanglement with atoms and photons \cite{Raimond01}, band gap engineering in condensed matter \cite{Kibis11}, nonlinear optical properties of nanostructures \cite{Savenko12}, currents induced by vacuum fluctuations \cite{Kibis13}, etc. ~--- together with the modern field of circuit QED \cite{Blais21}, is focused more on resonant interaction of atoms and nanostructures with single-mode fields in cavities or resonators. In contrast, waveguide QED describes the interaction of atom-like structures with a continuum of photon modes in open waveguides.

Three-level systems play an important role in quantum optics and photonics, and have recently received broad attention in waveguide QED as well. Interaction of three-level emitters (3LE) with photon fields can induce some interesting effects, such as electromagnetically induced transparency (EIT) \cite{Abdumalikov10}, Autler-Townes splitting \cite{Sillanpaa09}, cross-over of these effects \cite{Anisimov11}, two-photon resonance fluorescence from experimental \cite{Gasparinetti19} and theoretical \cite{Ngaha25} point of view, giant cross-Kerr effect \cite{Hoi13a}, photon-state purification \cite{Lopes24}, and many others. 

Since an excited three-level system can emit or absorb two photons, it can be potentially used to induce photon-photon correlations. They can be implemented via scattering two-photon pulses on a lambda-type \cite{Roy11} or ladder-type \cite{Li15} three-level system, although a four-level system also can be used to create correlated and entangled photon pairs \cite{Zheng12}. Creation of many-photon entangled states via interaction with a $\Lambda$-type atom was studied in \cite{Ilin24}. The other way to induce photon-photon correlations is to utilize the spontaneous emission process. It can be done by driving a three-level ladder-type atom with two laser fields for each transition \cite{Chen12}, or by using a single drive on the frequency of the 0-2 transition \cite{Sathyamoorthy16}. It is also possible to create a correlated photon pair with almost instant decay of an excited 3LE for a superconducting giant-atom \cite{Gao24}, although in typical experiments for artificial atoms the process of spontaneous emission is usually of a cascaded nature \cite{Gasparinetti17}.

A typical realization of a 3LE interacting with a continuum of photons in 1D waveguides is superconducting artificial atoms \cite{Kjaergaard20}, directly or side-coupled to a microwave transmission line \cite{Krantz19}. The most common type of a three-level system in this playground is a transmon qubit \cite{Koch07}. A transmon has a ladder-type structure of levels, meaning that each level can interact only with neighbouring levels. Although transmons are usually used for quantum information applications as regular qubits, i.e., two-level systems, the third level can have an important practical role in waveguide QED setups, for example to create a single-photon router \cite{Hoi11}, to characterize decoherence rates \cite{Lu21}, or to extract main parameters of qubit in open waveguide \cite{Sultanov25}. Moreover, three-level transmon structures have wide quantum optics applications, see \cite{Hoi13b} and references from the previous paragraph.

In this work we study the spontaneous emission of an initially excited three-level transmon, coupled to a 1D open waveguide. More traditional tasks like interaction of monochromatic wave \cite{Shen05} or photon pulse \cite{Chen11} with single qubit, or description of interaction of qubit chains with plane wave via Lippmann-Schwinger equation \cite{Fang14} or non-hermitian Hamiltonian approach \cite{Greenberg15}, or even scattering of Gaussian impulse on single qubit \cite{Greenberg23}, two-qubit system \cite{Greenberg24}, and on qubit chain \cite{Liao15}, are usually performed in the single-photon limit. In our work, however, we employ the two-excitation subspace. Our approach is based on the real-space description of photon fields. We derive an analytical solution of the non-stationary Schr\"odinger equation and find the probability amplitudes for the complete wavefunction of the system. Using this solution, we analyze probabilities to detect certain states of the system, and the spectral density of output two-photon fields. 

For a weak radiative decay rate we find that each of the two photons has a distinct frequency, corresponding to the two transitions of the 3LE, which is in good agreement with experiments \cite{Gasparinetti17}. However, if the radiative decay is rather strong, we find that frequency correlation appears, which can result in the creation of two identical photons with equal energies. This correlation is more pronounced for low anharmonicity of the 3LE i.e., when the levels are almost equidistant. We show that the spectral density for identical photons have two maxima: on resonance with the lower two levels, or shifted by half the anharmonicity parameter. Since superconducting systems have an inherent ability to tune their internal parameters, this type of frequency correlation can be potentially used to design a source of identical photons in waveguide QED setups. 

The paper is organized as follows. In Section II we give a general description of the system under study and derive the main equations for the wavefunction's amplitudes. In Section III we derive a general analytical solution in real-space. In Section IV we examine the case of a fully-excited three-level atom, both in real-space and in frequency representations. The main results are placed in this section, including analysis of state-detection probabilities, two-photon emission spectra and creation of identical photons. The conclusion of our work, as well as future prospects, are reflected in Section V. Some mathematical calculations are included in the Appendix.

\section{Real-space description of the transmon-waveguide system}

\begin{figure}
\centering
\includegraphics[width=0.5\textwidth]{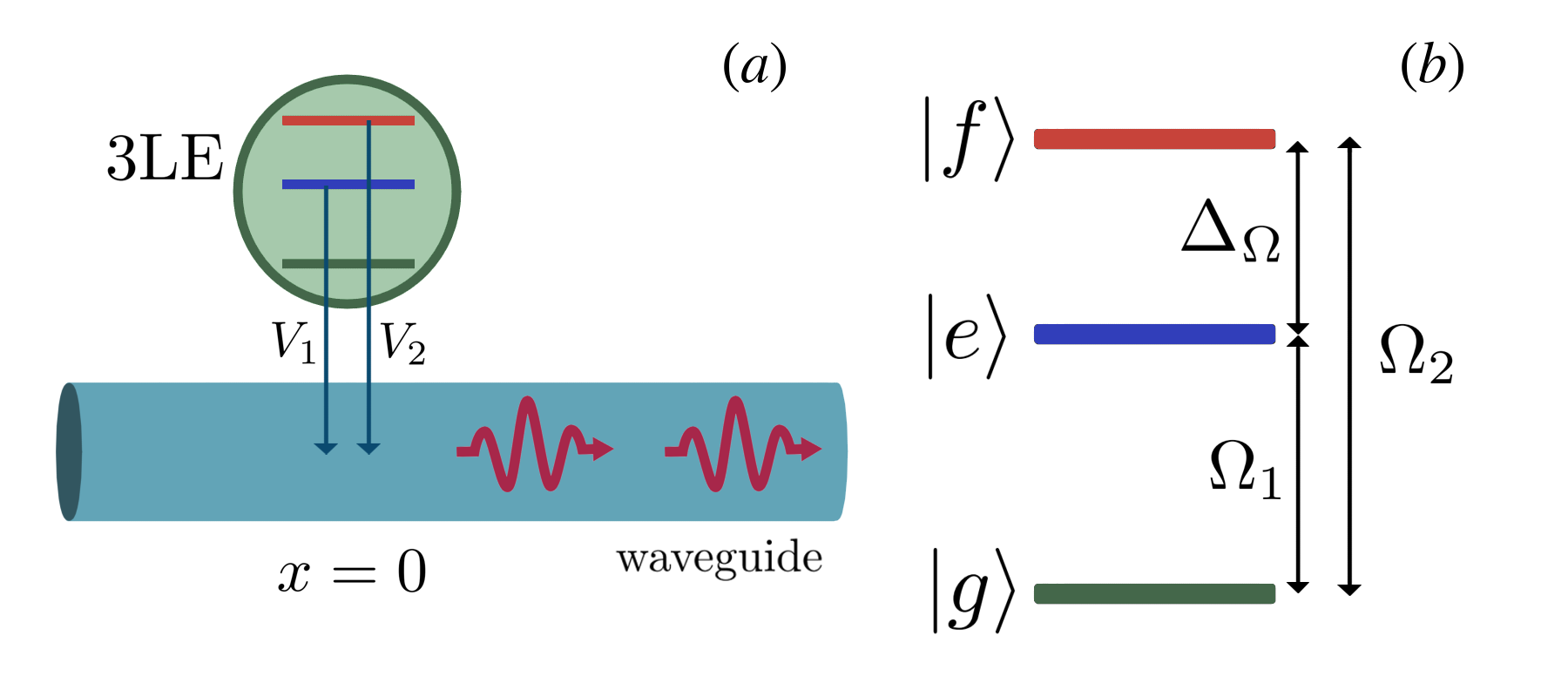}\\
  \caption{(a) System under study~--- a three-level emitter (3LE), coupled to the center of a one-dimensional waveguide at point $x=0$. The coupling of the first and second-excited states to the waveguide are $V_1$ and $V_2$, respectively. An excited atom can emit photons through spontaneous decay, which propagate only forward due to the usage of the chiral approximation (see text for details). (b) Energy levels diagram of the bare transmon and the corresponding resonant frequencies.}  \label{fig1} 
  \end{figure}

We consider a system of a ladder-type three-level emitter (3LE), located at the point $x=0$ in a one-dimensional waveguide (see figure \ref{fig1}(a)). In the real-space representation \cite{Shen07} and under the rotating wave approximation (RWA) it can be described by the following Hamiltonian (here and later in the text we assume that $\hbar = 1$):
  \begin{equation} \label{hamiltonian}
\begin{gathered}
H = \Omega_1 \left| e \right\rangle \left\langle e \right| + \Omega_2 \left| f \right\rangle \left\langle f \right| - i \upsilon_g \int\limits_{-\infty}^{+\infty} dx \; a^\dagger (x) \frac{\partial }{\partial x} a(x)
\\
+ \int\limits_{-\infty}^{+\infty} dx \; V_1 \delta (x) \left( { a^\dag (x) \left| g \right\rangle  \left\langle e \right| + a(x) \left| e \right\rangle \left\langle g \right| } \right)
\\
+ \int\limits_{-\infty }^{+\infty} dx \; V_2 \delta (x) \left( { a^\dag (x) \left| e \right\rangle \left\langle f \right| + a(x) \left| f \right\rangle \left\langle e \right|} \right).
\end{gathered} 
  \end{equation}
Here states $\left| g \right\rangle, \left| e \right\rangle$ and $\left| f \right\rangle$ are the ground, first-excited, and second-excited states of the transmon. The corresponding resonant frequencies of the first and second excited states are $\Omega_1, \Omega_2$, and $V_1, V_2$ are their coupling constants. Energy levels and corresponding frequencies are shown in figure \ref{fig1}(b). $\upsilon_g$ is the group velocity of photons, $\delta(x)$ is the Dirac delta-function, and $a^\dagger(x)$ ($ a(x)$) is the creation (annihilation) operator in the x-representation, which creates (destroys) a photon at position $x$ of the waveguide.

Throughout this paper we will refer to the 3LE as a transmon or an (artificial) atom, but keep in mind that our approach is universal and can be related to other physical realizations of three-level ladder-type systems.

The Hamiltonian (\ref{hamiltonian}) is written in the so-called continuous limit \cite{Blow90}, meaning that instead of a discrete summation over $x$ (or over modes $k$), we have a continuous integration. In this limit, the commutation relation between photon operators is as follows:
   \begin{equation}
\left[ a(x), a^\dagger(x') \right] = \delta(x -x'),
   \end{equation}
hence the photon operators, usually dimensionless for discrete modes, in the continuous limit have the dimension of the inverse square root of coordinate, $a(x) \sim 1/\sqrt{x}$. The dimension of the coupling constants $V_1, V_2$, correspondingly, is $V \sim \omega \sqrt{x}$.

The first two terms in (\ref{hamiltonian}) correspond to a free atom and a free field, and the following two terms describe the interaction of photon fields with the first and second excited states of the transmon, respectively. Note that there is no direct transition between the ground and the second-excited state $\left| f \right\rangle \leftrightarrow \left| g \right\rangle$. Hamiltonian (\ref{hamiltonian}) is practically the Jaynes-Cummings Hamiltonian, extended for a three-level system, and it conserves the number of excitations in the system. In this work, we adhere to two excitations due to the two possible transitions of the transmon. We should note that in real superconducting waveguides, there exist two dissipation sources which are usually described by the non-radiative decay rate $\Gamma_{nr}$ and the decay rate to non-guided modes $\Gamma_{ng}$. However, they are much less than $\Gamma_{1D}$, the decay rate to the waveguide guided mode. Therefore, the coupling efficiency $\beta=\Gamma_{1D}/(\Gamma_{1D}+\Gamma_{nr}+\Gamma_{ng})$ is close to unity. For superconducting systems it can reach values of $\beta \approx 0.999$ \cite{Sheremet23}, so our exclusion of dissipation is well justified.

Since our atom has three levels, the wave function should be written in the two-excitation subspace as follows:
   \begin{equation}
\begin{gathered}
\left| \Psi (t) \right\rangle  = \alpha (t) e^{-i\Omega_2 t} \left| f,0 \right\rangle + \int\limits_{-\infty}^{+\infty} dx e^{-i\Omega_1 t} \beta (x,t) a^\dagger (x) \left| e,0 \right\rangle 
\\
+ \frac{1}{\sqrt 2 } \iint\limits_{-\infty}^{+\infty} dx_1 dx_2 \; \gamma (x_1, x_2, t) a^\dagger (x_1) a^\dagger (x_2) \left| g,0 \right\rangle ,
\end{gathered}  \label{wavfunc}
   \end{equation}
where the first probability amplitude $\alpha(t)$ corresponds to a excited transmon in state $\left| f \right\rangle$ and no photons in the waveguide, the second amplitude $\beta(x,t)$ corresponds to a transmon excited to the $\left| e \right\rangle$ state and one photon at position $x$, and the final amplitude $\gamma(x_1, x_2, t)$ contains a non-excited atom and two photons at positions $x_1$ and $x_2$. State $\left| 0 \right\rangle$ designates the vacuum Fock state. The coefficient $1/\sqrt{2}$ in the last term is a normalization constant for a two-photon state in the Fock basis.

Because of the continuous limit \cite{Blow90}, probability amplitudes with photons have the dimension of the inverse square root of coordinate: $\beta(x,t) \sim 1/\sqrt{x}$, and $\gamma(x_1, x_2, t) \sim 1/x$ due to the presence of two photons.

We should also note that the wavefunction (\ref{wavfunc}), as well as the Hamiltonian (\ref{hamiltonian}), describe photon propagation only in the forward direction. This is the so-called chiral approximation, and it can be performed by separating the Hamiltonian for odd and even modes \cite{Shen07}. Chiral, or non-reciprocal, waveguides are realized by creating a suppression of photon propagation in the backward direction and can be implemented in current waveguide QED experimental setups \cite{Guimond20}. However, we are mainly interested in the case of spontaneous decay of an initially excited 3LE, which should be symmetrical with respect to the 3LE placement, and thus bidirectionality has little effect on our system. So, in order to simplify the analytical derivation, we choose to consider only forward propagation.

Using the standard non-stationary Schr\"odinger equation $i \partial_t \left| \Psi \right\rangle = H \left| \Psi \right\rangle$ with the Hamiltonian (\ref{hamiltonian}) and the wave function (\ref{wavfunc}), we derive the equations for all probability amplitudes:
   \begin{equation}
\frac{d}{dt} \alpha (t) =  - iV_2 {e^{i(\Omega_2 - \Omega_1) t} }\beta (0,t), \label{eq1}   
   \end{equation}
   \begin{equation}
  \begin{gathered}
\frac{d}{dt} \beta (x,t) + \upsilon_g \frac{\partial }{\partial x} \beta (x,t) =  - i V_2 \delta (x) {e^{i(\Omega_1 - \Omega_2) t} } \alpha (t) 
\\
- i\frac{2}{\sqrt 2 } V_1 {e^{i\Omega_1 t}} \gamma (x,0,t),  \label{eq2}
  \end{gathered}  
   \end{equation}
   \begin{equation}
\begin{gathered}
\frac{d}{dt} \gamma (x_1, x_2, t) + \upsilon_g \frac{\partial }{\partial x_1} \gamma (x_1, x_2, t) + \upsilon_g \frac{\partial }{\partial x_2} \gamma (x_1, x_2, t) 
\\
 =  - \frac{i}{\sqrt 2 } V_1 {e^{-i\Omega_1 t}} \left[ { \delta (x_1) \beta (x_2, t) + \delta (x_2) \beta (x_1, t)} \right].  \label{eq3}
\end{gathered}   
   \end{equation}
In the last two equations, we use the fact that photons are fundamentally indistinguishable, which in our case can be stated as $\gamma (x_1, x_2, t) = \gamma(x_2,x_1,t)$. All further calculations are based on equations (\ref{eq1}-\ref{eq3}).

\section{Probability amplitudes in real-space} 

To find explicit expressions for all three amplitudes of the wave function, we use a method incorporated in \cite{Mukhopadhyay24} for the calculation of quantum Rabi oscillation dynamics and also employed in \cite{Lopes24} for photon state purification via 3LE interaction. We have already used this approach in \cite{Chuikin24} for the description of stimulated emission in waveguide QED.

First, let's define a Fourier transform for amplitudes with a photonic part, which allows us to switch from real-space to k-space of frequency modes:
  \begin{subequations}
\begin{gather}
\beta (k,t) = \frac{1}{\sqrt{2\pi} } \int\limits_{-\infty}^{+\infty } {dx\;} {e^{-ikx}}\beta (x,t),
\\
\gamma (k_1, k_2, t) = \frac{1}{2\pi} \iint\limits_{-\infty }^{+\infty} {dx_1 dx_2 \; } {e^{-i k_1 x_1}} {e^{ik_2 x_2}} \gamma (x_1, x_2, t).
\end{gather}  \label{fourier}  
  \end{subequations}
Inverse transformation is done by changing the sign of the exponents.

Here and further we assume that the dispersion is linear, $k = \omega / \upsilon_g$, which is well justified if the atom and photon resonant frequencies are far from the cut-off frequency of the waveguide \cite{Shen09}.

By using (\ref{fourier}) in (\ref{eq3}) together with the integral representation of the delta function as $\delta(x) = 1/2\pi \int dk \; e^{ikx}$, we obtain an equation for the two-photon amplitude in k-space:
   \begin{equation}
  \begin{gathered}
\frac{d}{dt} \gamma (k_1, k_2, t) + i \upsilon_g (k_1 + k_2) \gamma (k_1, k_2, t) = 
\\
= - \frac{i}{ \sqrt 2 \sqrt {2\pi} } V_1 {e^{ - i\Omega_1 t}} \left[ {\beta (k_2, t) + \beta (k_1 ,t)} \right].
  \end{gathered} 
   \end{equation}
Formal solution of this equation is:
   \begin{equation}
  \begin{gathered}
\gamma (k_1, k_2, t) = -\frac{i}{ \sqrt 2 \sqrt {2\pi } } V_1 \times
\\
\int\limits_0^t d\tau {e^{-i\upsilon_g(k_1 + k_2)(t - \tau )}} {e^{-i\Omega_1 \tau }} \left[ {\beta (k_2, \tau ) + \beta (k_1, \tau )} \right],
  \end{gathered}
   \end{equation}
where we assume that at the initial time $t=0$ no two-photon states are present. Transforming this expression back to real-space using the inverse of (\ref{fourier}), we get:
   \begin{equation}
\begin{gathered}
\gamma (x_1, x_2, t) =  
\\
- i\frac{V_1}{\sqrt 2 \upsilon_g} {e^{-i\Omega_1 (t - \frac{x_1}{\upsilon_g} ) }} \beta \left( {x_2 - x_1,t - \frac{x_1}{\upsilon_g}} \right)\theta (x_1) \theta \left( t - \frac{x_1}{\upsilon_g} \right)
\\
 - i\frac{V_1}{\sqrt 2 \upsilon_g} {e^{-i\Omega_1(t - \frac{x_2}{\upsilon_g})}} \beta \left( {x_1 - x_2,t - \frac{x_2}{\upsilon_g}} \right) \theta (x_2) \theta \left( {t - \frac{x_2}{\upsilon_g}} \right),
\end{gathered}   \label{gamma}
   \end{equation}      
where we introduced a Heaviside step-function, $\theta(x>0) = 1,\; \theta(x<0) = 0$. In order to avoid a jump at $x=0$, we apply the standard regulation $\theta(0) = 1/2$.

The next thing we need is to find $\beta(x,t)$, which then can be used to determine the two-photon amplitude (\ref{gamma}). Taking $\gamma(x_1,x_2,t)$ for $x_2 = 0$ and substituting it into (\ref{eq2}), we obtain:
   \begin{equation}
  \begin{gathered}
\left( {\frac{\partial }{\partial t} + {\upsilon _g}\frac{\partial }{{\partial x}}} \right) \beta (x,t) =  - i V_2 \delta (x) {e^{i(\Omega_1 - \Omega_2) t}} \alpha (t) 
\\
- \frac{\Gamma_1}{2} \beta (x,t) - \Gamma_1 F(x,t),
  \end{gathered}  \label{11}
   \end{equation}
where we define the decay rate of the first-excited level as $\Gamma_1 = V_1^2/\upsilon_g$, as well as some additional parameter:
   \begin{equation}
F(x,t) = e^{i\Omega_1 \frac{x}{\upsilon_g}} \beta \left({ -x, t - \frac{x}{\upsilon_g}} \right) \theta (x) \theta \left( t - \frac{x}{\upsilon_g} \right).  \label{F}
   \end{equation}

To find an explicit expression for $\beta(x,t)$, we use the same procedure as for $\gamma(x_1,x_2,t)$. Multiplying (\ref{11}) by $e^{-ikx}/\sqrt{2\pi} $ and integrating over $x$, we can use the Fourier transform (\ref{fourier}) to rewrite equation (\ref{11}) in k-space. Then the formal solution can be found as:
   \begin{equation}
\begin{gathered}
\beta (k,t) = \beta_0(k) {e^{-(ik \upsilon_g + \Gamma_1 /2 )t}} 
\\
- i \frac{V_2}{\sqrt{2\pi} } \int\limits_0^t d\tau {e^{-(ik \upsilon_g + \Gamma_1/2)(t - \tau )}} {e^{-i(\Omega_2 - \Omega_1) \tau }} \alpha (\tau)
\\
- \Gamma_1 \int\limits_0^t d\tau {e^{-(ik \upsilon_g + \Gamma_1/2)(t - \tau )}}F(k, \tau ),
\end{gathered}  \label{13}
   \end{equation}
where $\beta_0(k)=\beta(k,0)$ is an amplitude for an incident photon with frequency $\omega= k \upsilon_g$ at the initial time $t=0$ and $F(k,t)$ is the Fourier transform of coefficient (\ref{F}). Translating (\ref{13}) back to real-space, we get:

\begin{widetext}

   \begin{equation}
  \begin{gathered}
\beta (x,t) = e^{-\frac{\Gamma_1}{2} t} \beta_0(x - \upsilon_g t)
- i\frac{V_2}{\upsilon_g} e^{-i \Delta_\Omega (t - \frac{x}{\upsilon_g} )} e^{- \frac{\Gamma_1 x}{2\upsilon_g} } \theta(x) \theta \left( {t - \frac{x}{\upsilon_g}} \right) \alpha \left( t - \frac{x}{\upsilon_g} \right)
\\
- \Gamma_1 \theta(x) \theta \left( t - \frac{x}{\upsilon_g}  \right)\int\limits_0^{x/\upsilon_g} dt' {e^{-\frac{\Gamma_1}{2} t' }} {e^{-i\Omega_1 \left( {t' - \frac{x}{\upsilon_g}} \right)}} \beta \left( {{\upsilon _g}t' - x, t - \frac{x}{\upsilon_g}} \right),
   \end{gathered}   \label{14}  
   \end{equation}
where we introduce the frequency between levels $\left| e \right\rangle$ and $\left| f \right\rangle$ as $\Delta_\Omega = \Omega_2 - \Omega_1$ (see figure \ref{fig1}(b)). The change of the upper limit in the last integral is due to the step function $\theta \left( x - \upsilon_g t' \right)$.

Expression (\ref{14}), aside from the amplitude $\alpha(t-x/\upsilon_g)$, also contains the shifted value $\beta(\upsilon_g t' -x, t- x/\upsilon_g)$, so it can't be considered explicit. However, by performing a corresponding shift in (\ref{14}), one can show that the recursive part will be cut off due to the Heaviside functions, and the remaining part will contain only terms proportional to the incident single-photon field $\beta_0$ and the amplitude of the second-excited state $\alpha(t)$. Detailed calculations can be found in Appendix A. Thus, we obtain the explicit expression for the single-photon amplitude:
   \begin{equation}
  \begin{gathered}
\beta (x,t) = e^{-\frac{\Gamma_1}{2} t} \beta_0(x - \upsilon_g t) - i\frac{V_2}{\upsilon_g} e^{ -i{\Delta_\Omega }(t - \frac{x}{\upsilon _g}) } e^{-\frac{{{\Gamma_1}x}}{2\upsilon_g}} \theta (x) \theta \left( {t - \frac{x}{\upsilon_g}} \right)\alpha \left( {t - \frac{x}{\upsilon_g} } \right)
\\
- \Gamma_1 \theta(x) \theta \left( {t - \frac{x}{\upsilon_g}} \right) {e^{- \frac{\Gamma_1}{2} \left( {t - \frac{x}{\upsilon_g}} \right)}} \int\limits_0^{x/{\upsilon_g}} dt' {e^{-\frac{\Gamma_1}{2} t'}} {e^{-i\Omega_1 \left( {t' - \frac{x}{\upsilon_g}} \right)}} \beta_0 \left[ {{\upsilon_g}(t' - t)} \right].
  \end{gathered}   \label{beta}
   \end{equation}   

\end{widetext}

By taking (\ref{beta}) at point $x=0$, we can easily find the remaining amplitude of the second-excited state from (\ref{eq1}):
   \begin{equation}
\alpha (t) = \alpha (0) e^{ - \frac{\Gamma_2}{2} t} - i V_2 e^{-\frac{\Gamma_2}{2} t} \int\limits_0^t d\tau e^{\left( {i\Delta_\Omega - \frac{\Gamma_1}{2} + \frac{\Gamma_2}{2} } \right)\tau } \beta_0( -\upsilon_g \tau ),   \label{alpha}   
   \end{equation}
where $\alpha(0)$ corresponds to the initial state with a fully excited transmon, and $\Gamma_2 = V_2^2/\upsilon_g$ is the decay rate of the second-excited state, similar to $\Gamma_1$.

The obtained expressions (\ref{gamma}), (\ref{beta}) and (\ref{alpha}) are the key results of this section. Given a particular initial state, we can derive an explicit analytic expression for the full wave function (\ref{wavfunc}). These equations admit two possible initial conditions: 1) The transmon is fully excited in state $\left| f \right\rangle$ and there are no free photons in the waveguide; 2) The transmon is excited to state $\left| e \right\rangle$ and there is a single incident photon with an arbitrary line shape $\beta_0 \equiv \beta(k,0)$. The first condition describes spontaneous emission of a fully excited three-level system, and we analyze it in detail below. The second one describes the process of stimulated emission with an added third level, and we leave this task for future studies. It is also possible to solve this problem for an initial state with two incident photons scattering on a 3LE, but we omitted it to simplify the calculations.

\section{Spontaneous decay of the fully-excited transmon}

\subsection{Solution in real-space}

Let's assume that initially there are no photons in the waveguide, and a three-level atom is in the upper excited state $\left| f \right\rangle$:
   \begin{equation}
\alpha(0) = 1, \qquad \beta_0(x) = 0.  \label{init}  
   \end{equation}
In this case we have the process of spontaneous decay, which, due to the three levels, results in the emission of two photons.

Using the initial condition (\ref{init}) in (\ref{alpha}), (\ref{beta}) and (\ref{gamma}), we find the exact expressions for all three probability amplitudes of the wavefunction:
   \begin{subequations}
\begin{gather}
\alpha(t) = e^{- \frac{\Gamma_2}{2} t},
\\
  \begin{gathered}
\beta (x,t) = -i \sqrt{ \frac{\Gamma_2}{\upsilon_g}} e^{-\left( i{\Delta_\Omega } + \frac{\Gamma_2}{2} \right) (t - x/\upsilon_g) } e^{- \frac{\Gamma_1}{2} \frac{x}{\upsilon_g}} \times 
\\
\theta (x) \theta \left( {t - \frac{x}{\upsilon_g}} \right),
  \end{gathered}
\\
  \begin{gathered}
\gamma(x_1,x_2,t) = - \sqrt{ \frac{\Gamma_1 \Gamma_2}{2\upsilon_g^2}} \times
\\ 
\left({ e^{ -i\Omega_1 \left( t - \frac{x_1}{\upsilon_g} \right)} e^{ -\left( i \Delta_\Omega + \frac{\Gamma_2}{2} \right) \left( t - \frac{x_2}{\upsilon_g} \right) }  e^{- \frac{\Gamma_1}{2} \frac{x_2 - x_1}{\upsilon_g}} \Theta (x_1,x_2,t) }\right.
\\
\left.{  + {e^{ -i{\Omega_1}\left( {t - \frac{x_2}{\upsilon_g}} \right)}} e^{- \left( {i \Delta_\Omega + \frac{\Gamma_2}{2}} \right) \left( {t - \frac{x_1}{\upsilon_g} } \right)} e^{ - \frac{\Gamma_1}{2}\frac{x_1 - x_2}{\upsilon_g}}  \Theta (x_2,x_1,t) }\right),
\end{gathered}
\end{gather}  \label{amplitudes} 
   \end{subequations}
where the composite theta-function is:
   \begin{equation}
\Theta(x_1,x_2,t) = \theta (x_1) \theta (x_2 - x_1) \theta \left( {t - \frac{x_1}{\upsilon_g}} \right) \theta \left( {t - \frac{x_2}{\upsilon_g}} \right).
   \end{equation}
Working within the chiral approximation implies that all fields are right-moving. Consequently, spontaneous photon emission is restricted to the forward direction along the positive x-axis, $x_{1,2} \equiv x > 0$, since we place transmon in the center of the waveguide $x=0$.

\subsection{State detection probabilities}

  \begin{figure*}
\centering
\includegraphics[width=1\textwidth]{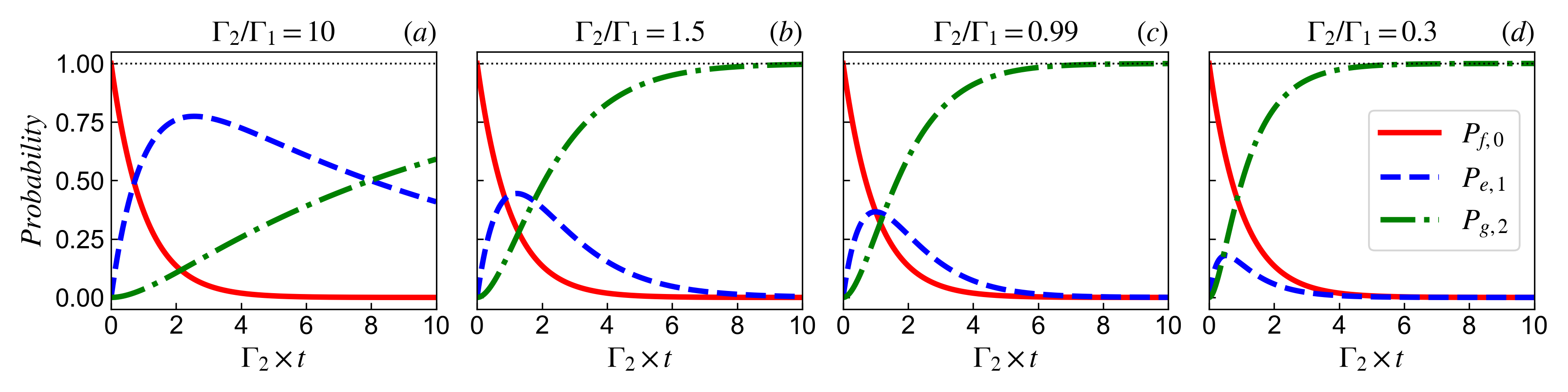}\\
  \caption{ Probabilities to detect a certain state in the entire waveguide, plotted with expressions (\ref{prob}): red solid line~--- transmon fully excited in level $\left| f \right\rangle$ and no photons present; blue dashed line~--- transmon in the first-excited state $\left| e \right\rangle$ and a single photon in the waveguide; green dash-dotted line~--- transmon in the ground state $\left| g \right\rangle$, and two photons in the waveguide. The sum of all probabilities is equal to one and shown by a thin dotted line. Different ratios of $\Gamma_2/\Gamma_1$ are shown above the plots. Time is measured in $\Gamma_2$ units. The decay rate of the upper level is constant, $\Gamma_2/\Omega_1 = 0.02$.}  \label{fig2} 
  \end{figure*}
  
Probability amplitudes (\ref{amplitudes}), besides $\alpha(t)$, contain a photonic dependence proportional to the real-space coordinate $x$ where the photon is presumably located. So we can treat the square root of those amplitudes as the density of probability to find a photon in a certain waveguide region. To find the full probabilities to locate a desired system state, one needs to integrate it over the entire waveguide. By doing so, we obtain:
   \begin{subequations}
\begin{gather}
P_{f,0}(t) = \left| \alpha(t) \right|^2 = e^{-\Gamma_2 t},
\\
P_{e,1}(t) = \int {dx} {\left| {\beta (x,t)} \right|^2} = \frac{\Gamma_2}{\Gamma_2 - \Gamma_1} \left( {e^{-\Gamma_1 t} - e^{-\Gamma_2 t}} \right),
\\
\begin{gathered}
P_{g,2}(t) = \iint dx_1 dx_2 \left| \gamma(x_1,x_2,t) \right|^2 =
\\
= \frac{\Gamma_1 \Gamma_2}{\Gamma_2 - \Gamma_1} \left( {\frac{1 - e^{ -\Gamma_1 t}}{\Gamma_1} - \frac{1 - e^{ -{\Gamma_2}t } }{\Gamma_2}} \right).
\end{gathered}
\end{gather}    \label{prob}
   \end{subequations}

As one can see from (\ref{prob}), the probabilities to detect a certain state of the wavefunction (\ref{wavfunc}) are independent of the atomic resonant frequencies and contain only the decay rates $\Gamma_1$ and $\Gamma_2$. Plots of state-detection probabilities for a few $\Gamma_2/\Gamma_1$ ratios are presented in figure \ref{fig2}.

The decay of the upper level $\left| f \right\rangle$ does not depend on $\Gamma_1$, i.e., it stays the same for different $\Gamma_2/\Gamma_1$ ratios since we choose to keep $\Gamma_2/\Omega_1 = 0.02$ constant (red solid line in figure \ref{fig2}). However, this is not true for the decay with the emission of a single or two photons in the entire waveguide. There are two extreme cases: the decay of the upper level is much faster than that of the middle one, $\Gamma_2 / \Gamma_1 \gg 1$, and the inverse picture, when $\Gamma_2 / \Gamma_1 \ll 1$. For the first case, the decay of the upper level is so fast that the transmon, after a short period of time, starts to behave somewhat similarly to a regular two-level system formed by states $\left| e \right\rangle$ and $\left| g \right\rangle$ (see blue and green lines in figure \ref{fig2}(a)). In the second case, the decay of the upper level $\left| f \right\rangle$ is almost instantly followed by the decay of the middle level $\left| e \right\rangle$, so, despite the cascaded process of emission, the two photons are produced nearly simultaneously (the blue line in figure \ref{fig2}(d) is very close to zero, and the green line is close to unity).

Typical transmon circuits have a coupling ratio between the first-excited and second-excited states of $g_2(k)/g_1(k) \sim \sqrt{3/2}$, see expressions (3.3-3.4) in \cite{Koch07}. The corresponding ratio of decay rates is thus $\Gamma_2/\Gamma_1 \sim 1.5$, although some works mention a very similar value of $\Gamma_2/\Gamma_1 \sim 2$ \cite{Gasparinetti17}. We plot the probabilities for the common relation of $\Gamma_2/\Gamma_1 = 1.5$ in figure \ref{fig2}(b). Albeit, recent progress in superconducting quantum circuits has made possible the creation of the so-called giant-atom transmon with a highly tunable ratio $\beta = \Gamma_2/\Gamma_1 \approx 0.3-60$ \cite{Vadiraj21}. However, further down we stick to more common transmon types with ratios of $\Gamma_2/\Gamma_1 = 1.5$ and $\Gamma_2/\Gamma_1 = 3$.

\subsection{Two-photon spectrum}

By using the Fourier transform (\ref{fourier}), we can translate the real-space photon amplitudes (\ref{gamma}) and (\ref{beta}) into k-space of frequency modes:
   \begin{equation}
\beta (k,t) = i\sqrt {\frac{\Gamma_2 \upsilon_g}{2\pi}} \frac{ e^{-\left( {i\omega + \frac{\Gamma_1}{2} } \right)t} - e^{ -\left( i \Delta_\Omega + \frac{\Gamma_2}{2}  \right) t } }{i(\omega - \Delta_\Omega ) - \frac{\Gamma_2 - \Gamma_1}{2} },  \label{beta_k}
   \end{equation}
   \begin{equation}
\begin{gathered}
\gamma (k_1, k_2,t) = \frac{\upsilon_g}{2\pi } \sqrt{ \frac{\Gamma_1 \Gamma _2 }{2} } \frac{ e^{-i(\omega_1 + \omega_2)t} }{ i(\omega_1 - \Delta_\Omega ) - \frac{\Gamma_2 - \Gamma_1}{2} } \times 
\\
\left[{ \frac{ e^{\left({ i\omega_2 - i\Omega_1 - \frac{\Gamma_1}{2} }\right)t} - 1 }{ i\omega_2 - i\Omega_1 - \frac{\Gamma_1}{2} } - \frac{{{e^{\left( { i\omega_1 + i\omega_2 - i\Omega_2 - \frac{\Gamma_2}{2} } \right)t}} - 1}}{ i\omega_1 + i\omega_2 - i\Omega_2 - \frac{\Gamma_2}{2} }  }\right]
\\
 + \frac{\upsilon_g}{2\pi} \sqrt{ \frac{\Gamma_1 \Gamma _2}{2} } \frac{ e^{-i(\omega_1 + \omega_2)t} }{i(\omega_2 - \Delta_\Omega ) - \frac{\Gamma_2 - \Gamma_1}{2} } \times 
\\
\left[{ \frac{{e^{\left({ i\omega_1 - i\Omega_1 - \frac{\Gamma_1}{2} }\right) t}} - 1}{i\omega_1 - i\Omega_1 - \frac{\Gamma_1}{2} } - \frac{{ e^{\left( {i\omega_1 + i\omega_2 - i\Omega_2 - \frac{\Gamma_2}{2} } \right)t}} - 1}{i\omega_1 + i\omega_2 - i\Omega_2 - \frac{\Gamma_2}{2} } }\right],
\end{gathered}   \label{gamma_k}
   \end{equation}
where in $\beta(k,t)$ the photon frequency $\omega$ corresponds to the first emitted photon.

Note that if we take $t \rightarrow \infty$, the single-photon amplitude (\ref{beta_k}), as well as $\alpha(t)$, will be equal to zero. However, the two-photon amplitude (\ref{gamma_k}) for large times takes a non-zero value. This is a logical prediction because the final state of an excited three-level system is obviously two photons in the waveguide with the transmon being in the ground state.

\begin{figure*}
\centering
\includegraphics[width=0.85\textwidth]{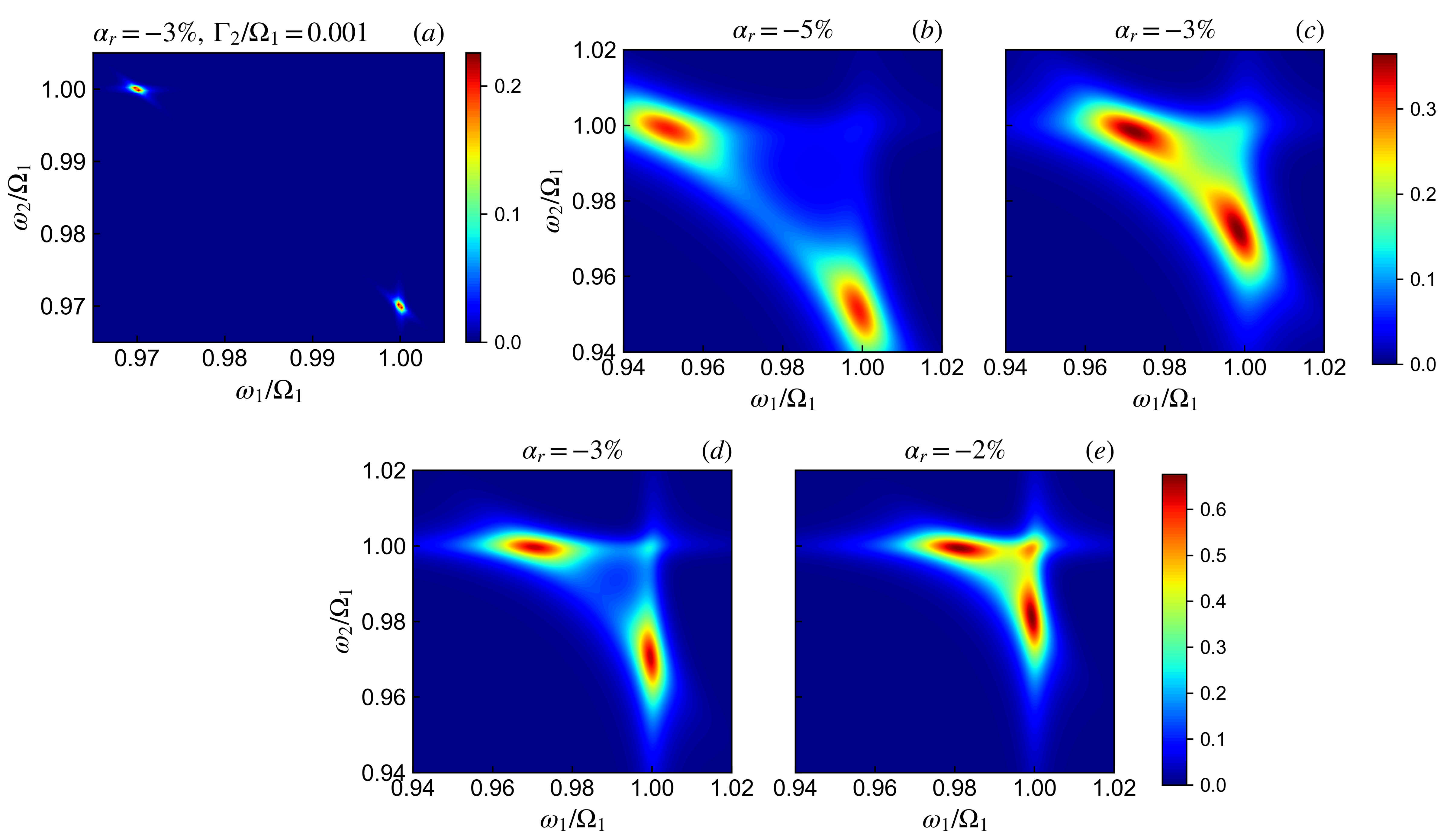}\\
  \caption{Two-photon spectral density of spontaneous emission from a three-level transmon $S_{sp}(\omega_1,\omega_2)\Gamma_2^2$, plotted using expression (\ref{spectra2}). (a) is plotted for weak-coupling regime $\Gamma_2 / \Omega_1 = 0.001$ and typical anharmonicity $\alpha_r = -3 \%$; Other plots describe stronger coupling for the constant decay rate $\Gamma_2 / \Omega_1 = 0.02$. (b) and (c) plotted for typical decay rates relation $\Gamma_2/ \Gamma_1 = 1.5$; (d) and (e) for faster decay of the upper level $\Gamma_2/ \Gamma_1 = 3$. Plots (b)-(c) share the same colorbar for a more direct comparison of spectral density values, the same applies to plots (d)-(e).}  \label{fig3} 
  \end{figure*}

Let's define the spectral density of spontaneous emission of an initially excited transmon as follows:
   \begin{equation}
\begin{gathered}
S_{sp}(\omega_1, \omega_2) = \upsilon_g^2 {\left| {\gamma (k_1,k_2,t \to \infty )} \right|^2} = 
\\
 = \frac{1}{8\pi^2} \frac{\Gamma_1 \Gamma_2}{ {{\left( \omega_1 + \omega_2 - \Omega_2 \right)}^2 + \frac{\Gamma_2^2}{4} } } \times 
\\
\frac{{{\left( \omega_1 + \omega_2 - 2\Omega_1 \right)}^2} + \Gamma_1^2 }{\left[ {{{\left( {\omega_1 - \Omega_1} \right)}^2} + \frac{\Gamma_1^2}{4} } \right] \left[{ {{\left( {\omega_2 - \Omega_1} \right)}^2} + \frac{\Gamma_1^2}{4}  }\right]}.
\end{gathered}   \label{spectra2}
   \end{equation}
We plot this two-photon spectral density in figure \ref{fig3}.
 
Unlike the probabilities (\ref{prob}), the spectral density (\ref{spectra2}) depends not only on the decay rates $\Gamma_1, \Gamma_2$, but also on the resonant frequencies of the 3LE. It is typical to describe the energy-level spacing of a transmon via an anharmonicity parameter \cite{Koch07}:
   \begin{equation}
\alpha_r = \frac{\Delta_\Omega - \Omega_1}{\Omega_1} = \frac{\Omega_2 - 2\Omega_1}{\Omega_1} \approx -{\left( \frac{8 E_J}{E_C} \right)^{-1/2}},  \label{anh}
   \end{equation}
where $E_J$ is the Josephson energy, and $E_C$ is the charging energy~--- common characteristics of superconducting quantum circuits \cite{Blais21}. The subscript $r$ stands for 'relative', thus expression (\ref{anh}) is dimensionless and shows the relation of energies between states $\left| f \right\rangle, \; \left| e \right\rangle$ and the energies between levels $\left| e \right\rangle, \; \left| g \right\rangle$. In literature it is also common to use a non-relative anharmonicity $\alpha = ( \Delta_\Omega - \Omega_1 ) = \alpha_r \Omega_1$, which is measured in hertz, but here we choose to use only the relative parameter to avoid confusion with the probability amplitude $\alpha(t)$. Typical experimental values of $\alpha$ usually fall within the range 100-300 MHz \cite{Krantz19}, or in relative units from $-2 \%$ up to $-6 \%$. Note that the anharmonicity for superconducting qubits is always negative, due to the smaller spacing between the upper levels.

From figure \ref{fig3}(a)-(b) we see that the spectral density of spontaneous emission has two distinct peaks, which are a result of the non-distinguishability of photons. Each peak corresponds to one resonant photon with frequency $\omega = \Omega_1$ and a second photon with a smaller frequency $\omega = (1+\alpha_r)\Omega_1$ (remember that $\alpha_r$ is negative). But because both photons are indistinguishable, we can't exactly say which photon is 'first' and which is 'second'; thus two peaks appear. For bigger anharmonicity, the two peaks are further apart because of the change of transition frequencies in the 3LE. If we implement weak coupling, $\Gamma/\Omega_1 = 0.001$ (in practical circuits it is usually $\Gamma < 5$ MHz), both peaks tighten into almost point-like shapes (see figure \ref{fig3}(a)). That means that we have two photons with distinct frequencies, corresponding to atom-transitions, $\Omega_1$ and $(1+\alpha_r)\Omega_1$. This behavior is in good agreement with the experimental work by Gasparinetti et al. \cite{Gasparinetti17}.

However, if we take stronger coupling, typical for circuit QED systems (for example, $\Gamma_2 \sim 100$ MHz), we see the appearance of frequency correlation between the two photons. This is expressed by the appearance of a third peak at the resonant frequency $\omega/\Omega_1 = 1$, which starts to become visible for $\Gamma_2/\Gamma_1 = 1.5$ in figure \ref{fig3}(c), and is more pronounced for $\Gamma_2/\Gamma_1=3$ in figure \ref{fig3}(d)-(e). Thus, even if the three-level system has a non-symmetrical level structure, there is a probability that during spontaneous decay the emitted photons will have the same frequency $\omega_1 = \omega_2 = \Omega_1$. This effect is connected to interference between the first emitted photon and the 3LE being still excited in state $\left| e \right\rangle$, which arises due to the almost equidistant level structure of a low-anharmonicity transmon and rather strong coupling between the artificial atom and the field modes.

The observation of a frequency-correlated photon pair naturally raises the question of whether the two emitted photons are entangled. Entanglement of two photons requires some kind of distinction between different photon states, for example in two polarizations, that can be encoded as two different photon modes. In our work we have only a right moving photon mode, described by photon creation/annihilation operators $a^\dagger(x)/ a(x)$. Thus, in our derivation we cannot investigate the entanglement of the emitted photon pair properly. However, by considering a bidirectional waveguide with forward-propagating mode $a^\dagger(x)$ and backward-propagating mode $b^\dagger(x)$ one could investigate the spatial entanglement of two photons, for example in the so-called N00N state $\left( \left|2_R,0_L \right\rangle - \left|0_R,2_L \right\rangle \right)/\sqrt{2}$. We believe that this kind of problem is interesting in itself and warrants its own dedicated research.

\subsection{Emission spectra for identical photons} 

  \begin{figure*}
\centering
\includegraphics[width=\textwidth]{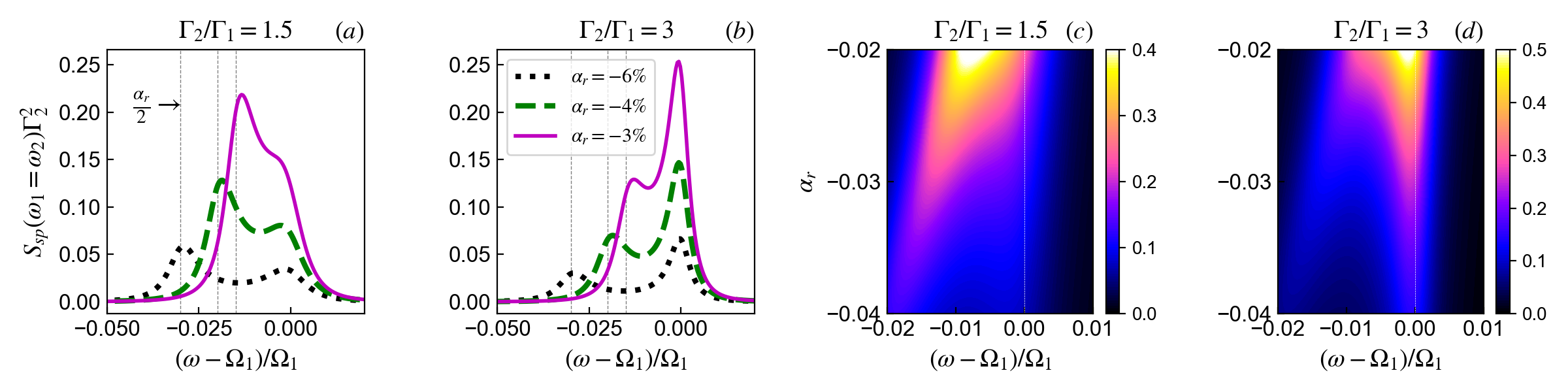}\\
  \caption{Spectral density of two identical photons (\ref{spectra1}) vs. frequency detuning~--- (a) and (b), and vs. detuning and anharmonicity~--- (c) and (d). (a) and (c) are plotted for the $\Gamma_2/\Gamma_1 = 1.5$ ratio, and (b) and (d) for $\Gamma_2/\Gamma_1 = 3$. In (a) and (b), the solid magenta line corresponds to $\alpha_r=-3 \%$, the green dashed line to $\alpha_r = -4 \%$, and the black dotted line to $\alpha_r=-6 \%$. Vertical grey lines show the points for $\delta_\omega/\Omega_1 = \alpha_r/2$. In (c) and (d), thin white lines indicate the resonant frequency $\omega=\Omega_1$. The decay of the upper level is taken constant $\Gamma_2 / \Omega_1 = 0.02$.}  \label{fig4} 
  \end{figure*}
   
Let's study the spectral density (\ref{spectra2}) for identical photons in closer detail. Taking $\omega_1 = \omega_2$ we get:
   \begin{equation}
S_{sp}(\omega_1=\omega_2) = \frac{1}{2 \pi^2} \frac{\Gamma_1}{\delta_\omega^2 + \frac{\Gamma_1^2}{4}} \times \frac{\Gamma_2}{ (2\delta_\omega - \alpha_r \Omega_1)^2 +\frac{\Gamma_2^2}{4}},   \label{spectra1}
   \end{equation}
where we introduce the detuning from the resonant frequency for both photons as $\delta_\omega = \omega - \Omega_1$. We can interpret function (\ref{spectra1}) as a diagonal cross-section from two-photon plots from figure \ref{fig3}.
   
The function (\ref{spectra1}) has two maxima: the first one corresponds to the frequency between the lower levels $\Omega_1$ and thus $\delta_\omega = 0$, and the second one corresponds to the decay of the upper level with an anharmonicity shift, $\delta_\omega = - \alpha_r \Omega_1 /2$. Both these peaks can be observed in figure \ref{fig4}(a)-(b), for the decay rates $\Gamma_2/\Gamma_1=1.5$ and $\Gamma_2/\Gamma_1=3$. A two-dimensional plot of (\ref{spectra1}), showing how the spectral density also changes with anharmonicity, is presented in figure \ref{fig4}(c)-(d).

The maximum of the spectral density for identical photons should be for $\alpha_r \rightarrow 0$, i.e., in the case of equidistant levels with $\Omega_2 = 2 \Omega_1$. However, as can be seen from figure \ref{fig4}, the frequencies of the two photons do not always have to be resonant with $\Omega_1$. For example, if the ratio between the decay rates is $\Gamma_2/\Gamma_1 = 1.5$, like in figure \ref{fig4}(a), the emitted photons both would have the maximum spectral density at a frequency close to $\Omega_1(1+\alpha_r/2)$. Thus, by changing both the decay rates $\Gamma_1, \Gamma_2$ and the anharmonicity parameter $\alpha_r$ of the excited three-level transmon, it is possible to create correlated photon pairs with the same frequencies. Superconducting circuits are very well suited for this task, since they allow tuning of the couplings and anharmonicity parameters either by sample production techniques or via the application of external magnetic fields.

\section{Conclusion}

In summary, we present a theoretical description of spontaneous decay of a fully-excited three-level system in a one-dimensional open waveguide. Using the real-space description of the system, we obtain non-stationary analytical expressions for the probability amplitudes for the whole system. We show that the dynamics of spontaneous emission is heavily dependent on the ratio $\Gamma_2/\Gamma_1$, where $\Gamma_2$ ($\Gamma_1$) is the radiative decay rate of the second (first) excited state. Transferring the obtained amplitudes into the momentum space allowed us to study the spectral density of emitted photons, from which we discover the appearance of frequency correlation for strong coupling. It is expressed in the creation of two photons with identical frequencies, even if the three-level atom structure is anharmonic. The frequency correlation is stronger for smaller anharmonicity, i.e., when the levels are close to equidistant.

More detailed analysis showed that the spectral density of identical photons can have two maxima. The first one corresponds to the resonant frequency of the lower levels $\omega = \Omega_1$ and prevails when the decay rates ratio is rather high, $\Gamma_2/\Gamma_1 \approx 3$. The second maximum, however, depends on the anharmonicity parameter as $\omega \approx \Omega_1(1+\alpha_r/2)$, and it dominates for $\Gamma_2/\Gamma_1 \approx 1.5$, which is a rather common value for typical transmon circuits. Thus, by adjusting these parameters one can obtain pairs of identical photons with tunable frequency, which can be utilized for quantum information and communication applications.

Throughout our analysis we take specific parameters that are common for superconducting transmon qubits, coupled to a microwave coplanar waveguide. However, our results are obtained in a general form, and can be used to describe other physical systems of three-level emitters in 1D waveguides, like Rydberg atoms or quantum dots in nanoplasmonic waveguides.

Here we describe only the process of spontaneous emission of a fully-excited three-level emitter. However, the obtained solutions in real-space (\ref{gamma}), (\ref{beta}) and (\ref{alpha}) allow one to study also the case of an incident photon pulse, scattering on a half-excited three-level system in state $\left| e \right\rangle$. Moreover, our expression (\ref{beta}) is valid for an arbitrary pulse shape $\beta_0(x)$, which can be, for example, gaussian or rectangular. This task is quite interesting from the point of view of stimulated emission, and can be addressed in future studies.

\begin{acknowledgments}

The authors thank A. N. Sultanov for fruitful discussions and helpful comments on the manuscript. The work is supported by the Ministry of Science and Higher Education of Russian Federation under the project FSUN-2026-0004. O. Chuikin and O. Kibis acknowledge the financial support from the Foundation for the Advancement of Theoretical Physics and Mathematics “BASIS”.

\end{acknowledgments}

\appendix

\begin{widetext}

\section{Derivation of single-photon amplitude (\ref{beta}) }
We start by incorporating the following shift:
   \begin{equation}
x \to {\upsilon_g}t' - x; \qquad t \to t - \frac{x}{\upsilon_g};
   \end{equation}
and using it in (\ref{14}). By doing so we obtain:
   \begin{equation}
\beta \left( {{\upsilon _g}t' - x,t - \frac{x}{{{\upsilon _g}}}} \right) = {A_0} - \theta ({\upsilon _g}t' - x)\theta (t - t')\left[ {{A_1} + {A_2}} \right],  \label{A2}   
   \end{equation}
where for simplicity we introduce some coefficients as follows:
   \begin{subequations}
\begin{gather}
A_0 = e^{ -\frac{\Gamma_1}{2} (t - x/\upsilon_g)} \beta_0 \left[ {{\upsilon_g}(t' - t)} \right],
\\
A_1 = {\Gamma_1} \int\limits_0^{t' - x/{\upsilon_g}} dt'' {e^{- \frac{\Gamma_1}{2} t''}} {e^{-i\Omega_1 \left( {t'' - t' + \frac{x}{\upsilon_g}} \right)} } \beta \left( {x,t - t'} \right),
\\
A_2 = i\frac{V_2}{\upsilon_g} {e^{-i{\Delta_\Omega }(t - x/{\upsilon_g})} }{e^{(i{\Delta_\Omega } - {\Gamma_1}/2) (t' - x/{\upsilon_g})} } \alpha \left( {t - t'} \right).
\end{gather}   
   \end{subequations} 

Substituting (\ref{A2}) back to (\ref{14}), we get:
   \begin{equation}
\begin{gathered}
\beta (x,t) = {e^{ -\frac{\Gamma_1}{2} t}}{\beta_0}(x - {\upsilon_g} t) - i\frac{V_2}{\upsilon_g} {e^{ -i{\Delta_\Omega }(t - x/{\upsilon_g})}} {e^{ -\frac{{\Gamma_1}x}{2\upsilon_g}}} \theta (x) \theta \left( {t - \frac{x}{\upsilon_g}} \right)\alpha \left( {t - \frac{x}{\upsilon_g}} \right)
\\
 - {\Gamma _1}\theta (x)\theta \left( {t - \frac{x}{\upsilon_g}} \right)\int\limits_0^{x/{\upsilon_g}} {dt'} {e^{ -\frac{\Gamma_1}{2}t'}} {e^{ -i{\Omega_1}\left( {t' - \frac{x}{\upsilon_g}} \right)}}\left( {{A_0} - \theta ({\upsilon _g}t' - x)\theta (t - t')\left[ {{A_1} + {A_2}} \right]} \right).
\end{gathered} \label{A4} 
   \end{equation} 
The terms, proportional to $\int_0^{x/\upsilon_g} {\theta ({\upsilon_g}t' - x)dt'}$ will drop out due to mismatch of intervals, since the condition of the theta function is $t’ > x/\upsilon_g$. Thus, only term containing $A^{(0)}$ remain, proportional to the initial field $\beta_0$. Rewriting (\ref{A4}) in explicit form, we get the final expression (\ref{beta}).

\end{widetext}

\end{document}